\documentclass[]{spie}  

\newcommand{\LO}{$\mathcal{L}_{0}$}

\newcommand{\tablefoot}[1]{%
  \par\medskip
  {\footnotesize #1\par}
}

\usepackage{amsmath,amsfonts,amssymb}
\usepackage{graphicx}
\usepackage[colorlinks=true, allcolors=blue]{hyperref}

\title{Focal-Plane Diagnostics of Atmospheric and Dome-Induced Turbulence: Exploring techniques and applications}

\author[a,b]{Begoña García-Lorenzo}
\author[c,b]{Seifeldin Abouelella}
\author[c,b]{Kabir Baig}
\author[b]{Zenaida García Rodríguez}
\author[c,b]{Kshitij Ghimire}
\author[b]{Moisés Pulido Torres}
\author[b]{Alejandro Trujillo Ramos}
\author[a,b]{Jose A. Acosta-Pulido}
\author[a,b]{Julio A. Castro-Almazán}
\author[d,a,b]{Donaji Esparza-Arredondo}
\author[b,a]{Antonio Eff-Darwich}
\affil[a]{Instituto de Astrofísica de Canarias, C/ Vía Láctea s/n, 38205 La Laguna, Tenerife, Spain}
\affil[b]{Universidad de La Laguna, 38200 La Laguna, Tenerife, Spain}
\affil[c]{Rhine-Waal University of Applied Sciences, Kleve, Germany}
\affil[d]{Instituto de Radioastronomía y Astrofísica (IRyA), Universidad Nacional Autónoma de México, Antigua Carretera a Pátzcuaro \#8701, C.P. 58089 Morelia, Michoacán, México}
\authorinfo{Further author information: (Send correspondence to B.G-L.)\\B.G-L.: E-mail: begona.garcia@iac.es}

\begin{document} 
\maketitle

\begin{abstract}
Atmospheric turbulence remains the dominant source of image degradation in ground-based astronomy. Its impact depends not only on the integrated turbulence strength but also on the wavefront spatial-coherence outer scale (\LO) and on turbulence generated within the telescope enclosure (dome seeing). While seeing, coherence time, and related parameters are routinely monitored at major observatories, measurements of \LO\ and dome-induced turbulence remain sparse, limiting our ability to predict image quality and to construct accurate point-spread function (PSF) models.
Building on methodology recently developed for seeing-limited integral-field spectroscopic (IFS) data, we present a series of complementary studies showing how spatial information encoded in standard observations can be used to diagnose both atmospheric and dome turbulence. We investigate the feasibility of extending these techniques beyond IFS observations to other seeing-limited instruments through the analysis of wavelength-dependent spatial profiles. We also present the development of data-driven three-dimensional PSF models based on atmospheric statistics for the reconstruction of seeing-limited IFS observations.
These results show that widely available seeing-limited scientific observations, both from archival data and routine observatory operations, can provide meaningful and operationally useful constraints on key turbulence parameters and on the spectral behavior of dome seeing, complementing dedicated site-testing instrumentation. The resulting turbulence diagnostics have direct applications to PSF modelling, image reconstruction, image-quality prediction, and the optimization of future ELT operations.
\end{abstract}


\section{INTRODUCTION}
\label{sec:intro}  

Atmospheric turbulence strongly limits the angular resolution achievable by ground-based optical and infrared telescopes by distorting the incoming wavefront and degrading image quality (IQ). Although adaptive optics (AO) systems compensate for part of the atmospheric wavefront perturbations, the final IQ still depends on the statistical properties of atmospheric turbulence and on additional turbulence generated locally within telescope enclosures. The increasing aperture sizes of telescopes and the advent of Extremely Large Telescopes (ELTs) have reinforced the need for a more complete statistical characterization of turbulence parameters beyond traditional seeing monitoring.

The characterization of atmospheric turbulence is essential for the development of efficient AO systems, accurate point-spread function (PSF) reconstruction, and the optimization of observing strategies under prevailing atmospheric conditions. The turbulence parameters most relevant for these applications include the seeing ($\epsilon_{0}$), the coherence time, the isoplanatic angle, and the wavefront spatial-coherence outer scale (\LO). While the first three parameters are typically well characterized at major observatories, \LO\ remains comparatively poorly constrained and is generally derived only through dedicated or sporadic observing campaigns. Although current operational AO systems can provide estimates of \LO\ from telemetry data, they usually operate under favourable atmospheric conditions and still show significant variability in their sensitivity to \LO\ [\citenum{2023Barrientos}]. 

Existing techniques for estimating \LO\ often require dedicated instrumentation or AO telemetry analysis, limiting their operational applicability. As a consequence, IQ prediction and PSF reconstruction frequently rely on simplified assumptions regarding turbulence statistics and enclosure conditions. Recently, [\citenum{2024Garcia-Lorenzo}] proposed and empirically validated a methodology to estimate atmospheric turbulence parameters from seeing-limited integral-field spectroscopic (IFS) observations. The method exploits the intrinsic homogeneity of IFS data cubes to generate narrow-band images over a broad wavelength range obtained under common atmospheric and instrumental conditions. The wavelength dependence of the IQ measured from these long-exposure images can then be compared with analytical predictions derived from von Kármán turbulence statistics [\citenum{2002Tokovinin}], allowing the prevailing seeing and atmospheric outer scale during the observations to be constrained. 

Although typically weaker than atmospheric turbulence, another potential contributor to image degradation is the so-called dome seeing ($\epsilon_{\mathrm{dome}}$), arising from turbulence generated within telescope enclosures. Large telescope domes and their operational procedures are designed to minimize dome-induced turbulence; nevertheless, local turbulence can still contribute significantly to image degradation, not only through its strength but also through departures from the standard Kolmogorov turbulence regime. Despite its potential importance for IQ and PSF modelling, $\epsilon_{\mathrm{dome}}$ remains poorly characterized, particularly regarding its wavelength dependence. The method introduced in [\citenum{2024Garcia-Lorenzo}] also provides sensitivity to dome-induced turbulence whose wavelength dependence may depart from the Kolmogorov slope.

In this contribution, we build on [\citenum{2024Garcia-Lorenzo}] and explore how the same principle can be extended beyond its original IFS implementation. The underlying requirement is not the use of IFS itself, but the availability of spatial information at multiple wavelengths obtained under homogeneous observing conditions. We therefore investigate the potential application of this approach to other seeing-limited observing modes, and discuss its use as a focal-plane diagnostic of atmospheric and dome-induced turbulence. In addition, we explore how these diagnostics can provide physically motivated constraints for data-driven three-dimensional PSF models in seeing-limited observations.

These developments broaden the use of routine scientific observations for turbulence characterization and have direct applications to PSF reconstruction, IQ prediction, observing-strategy optimization, and future ELT operations.

\section{Spectroscopic observations as turbulence diagnostics}
\label{sec:ifs}

Building on the methodology introduced by [\citenum{2024Garcia-Lorenzo}], seeing-limited IFS observations can be used to constrain atmospheric turbulence parameters directly from standard scientific data. The approach exploits the simultaneous spatial and spectral information provided by IFS data cubes to generate multiple narrow-band images spanning a broad wavelength range under identical atmospheric and instrumental conditions. Measuring the wavelength dependence of the IQ in these long-exposure images allows direct comparison with analytical predictions derived from von Kármán turbulence statistics [\citenum{2002Tokovinin}],

\begin{equation}
\mathrm{IQ}_{LE}(\lambda) \approx \epsilon_0(\lambda)\sqrt{1-2.183\left(\frac{r_0(\lambda)}{\mathcal{L}_{0}}\right)^{0.356}}
\label{eq:Tokovinin}
\end{equation}

where IQ$_{LE}$ is the IQ of the long-exposure image, and $\epsilon_{0}$, $r_{0}$, and $\mathcal{L}_{0}$ denote the atmospheric seeing, Fried parameter, and atmospheric outer scale prevailing during the exposure integration time, respectively. 

The empirical analysis presented in [\citenum{2024Garcia-Lorenzo}] showed that the measured chromatic behavior of the IQ is sensitive not only to the atmospheric seeing and outer scale, but also to additional focal-plane broadening mechanisms such as dome-induced turbulence and possible departures from standard Kolmogorov turbulence statistics. Under the assumption of negligible instrumental aberrations, the total focal-plane IQ can be approximated as

\begin{equation}
\mathrm{IQ}_T^2(\lambda) \approx \mathrm{IQ}_{LE}^2(\lambda) + \epsilon_{dome}^2(\lambda),
\label{eq:IQ_total}
\end{equation}

where the dome-seeing contribution may follow a non-Kolmogorov wavelength dependence parameterized as

\begin{equation}
\epsilon_{dome}(\lambda) \propto \lambda^{(\gamma-4)/(\gamma-2)},
\label{eq:dome_lambda}
\end{equation}

with $\gamma = 11/3$ corresponding to the Kolmogorov regime. In this context, the key observable is the wavelength dependence of the IQ, quantified through the full width at half maximum (FWHM) of the PSF under homogeneous observing conditions.

\begin{figure} [ht]
   \begin{center}
   \begin{tabular}{c} 
   \includegraphics[width=14cm]{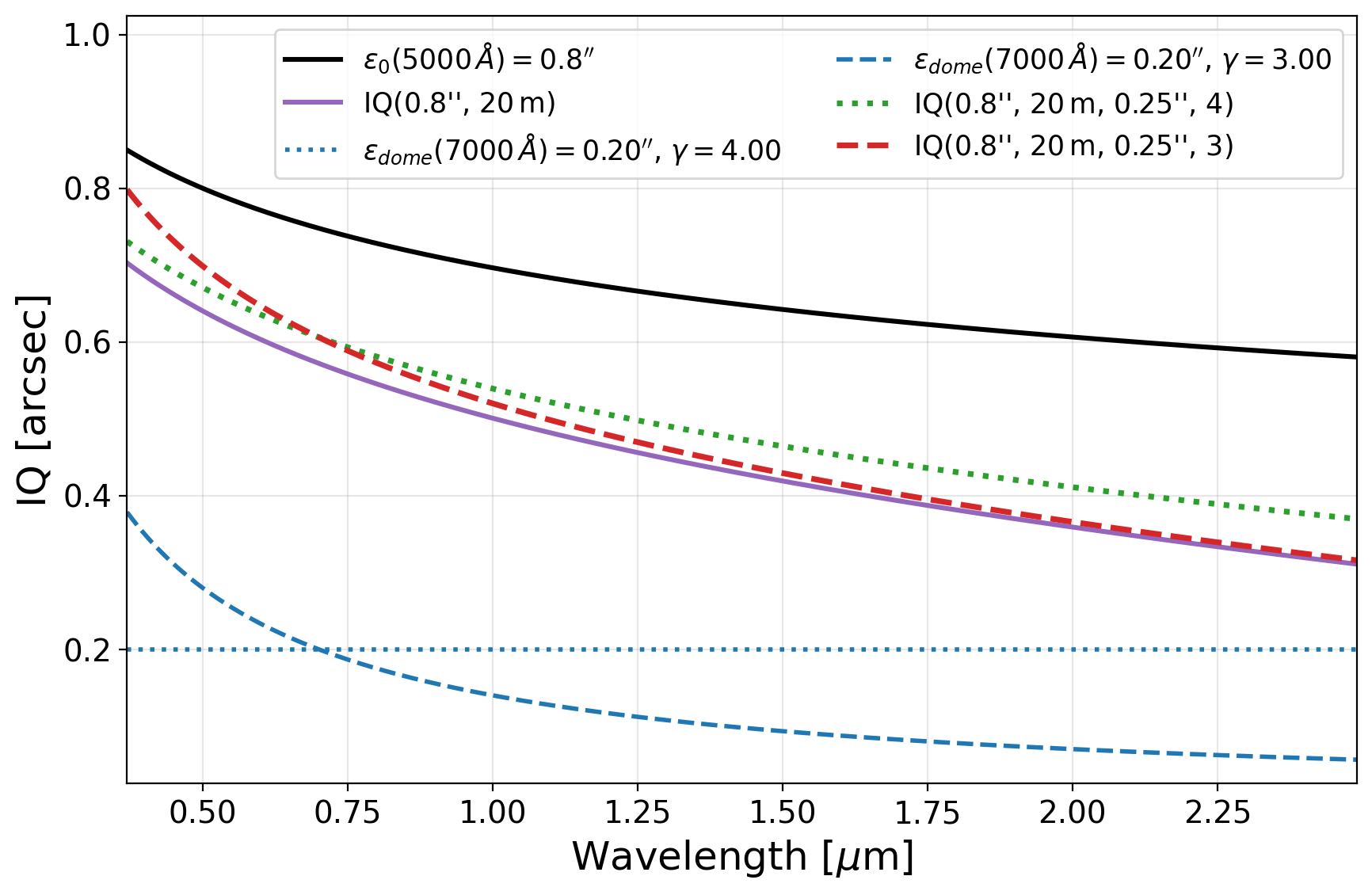}
   \end{tabular}
   \end{center}
   \caption[ ] 
   { \label{fig:Tokovinin} 
Wavelength dependence of the long-exposure IQ computed using the analytical approximation derived in [\citenum{2002Tokovinin}] (Eq.~\ref{eq:Tokovinin}) for an outer scale \LO = 20 meters and a seeing of $\epsilon_{0}(5000\,\AA)=0.8''$, representative of typical conditions at high-quality astronomical sites. The black solid line shows the Kolmogorov seeing scaling $\epsilon_{0}(\lambda)\propto\lambda^{-1/5}$, while the purple curve corresponds to the Eq.~\ref{eq:Tokovinin} prediction. 
Blue curves represent the dome-seeing contribution, assuming $\epsilon_{\mathrm{dome}}(7000\,$\AA$)=0.2''$ and spectral slopes (see Eq. \ref{eq:dome_lambda}) of $\gamma=4$ (dotted) and $\gamma=3$ (dashed). The green and red curves show the total $IQ_{T}(\lambda)$ obtained from the quadratic combination of atmospheric and dome contributions for the two dome-seeing models.
} 
   \end{figure}

Figure \ref{fig:Tokovinin} illustrates the predicted wavelength dependence of the IQ at the focal plane of a ground-based telescope under typical seeing-limited conditions. This figure highlights that dome turbulence modifies not only the absolute IQ but also its wavelength behaviour. The purple curve represents the atmospheric contribution computed from Eq. \ref{eq:Tokovinin} assuming a seeing of 0.8 arcsec and an atmospheric outer scale of 20 m, in the absence of dome-induced turbulence. The additional colored curves illustrate the effect of dome seeing with a strength of 0.2 arcsec under two extreme non-Kolmogorov turbulence regimes, highlighting how local turbulence can modify the wavelength behavior of the observed IQ (IQ$_T$). 

Although IFS provides an ideal implementation, the method only requires spatial information at multiple wavelengths obtained under common observing conditions. Long-slit spectroscopy, simultaneous multi-band imaging, and other seeing-limited observing modes may therefore provide equivalent turbulence diagnostics. This possibility motivates the exploratory analyses presented in the following sections.

\section{Extending turbulence diagnostics beyond IFS observations}
\label{sec:2}

The methodology briefly outlined above is based on measuring the wavelength dependence of the focal-plane PSF under homogeneous, seeing-limited observing conditions. Although IFS observations naturally provide simultaneous spatial and spectral information, similar diagnostics may also be possible with other observing techniques capable of delivering spatial profiles at multiple wavelengths. Extending these diagnostics beyond IFS observations is particularly attractive because of the large amount of archival data available and the possibility of implementing operational turbulence diagnostics without dedicated instrumentation. Nevertheless, the turbulence parameters derived from these observations must be validated against independent turbulence characterization measurements in order to assess their physical reliability and their correspondence with the actual atmospheric conditions before they can be used quantitatively.

\subsection{Long-slit spectroscopy}

Long-slit spectroscopy provides a natural extension of the methodology described above. In this case, the data can be interpreted as a simplified version of an IFS data cube, where the two-dimensional spatial information of an IFS observation, Nx,Ny,N$\lambda$, is reduced to a one-dimensional spatial sampling along the slit direction, that is, $\sim$Nx,1,N$\lambda$. Although these observations do not preserve the full two-dimensional PSF information available in IFS data cubes, they still retain the spatial profile of the source along the slit direction over a broad wavelength range under common atmospheric and instrumental conditions.

In this case, the observable is no longer the full focal-plane PSF, but its one-dimensional projection along the spatial axis of the slit. The wavelength dependence of this spatial profile can nevertheless be analysed in a manner analogous to the IFS case, allowing the extraction of chromatic IQ variations. For this application, the slit width should ideally be much larger than the prevailing seeing to minimise truncation of the spatial profile and to ensure adequate sampling of the seeing at the focal plane. We expect that slit widths of at least $\sim$3 times the seeing are likely sufficient to provide a reliable profile for these measurements, although this assumption should be assessed in detail in the context of a realistic implementation of the method. This condition is commonly fulfilled in observations of spectrophotometric standard stars obtained for flux-calibration purposes, where wide slits are routinely used.

As an exploratory test, we analysed archival long-slit spectroscopic observations of a spectrophotometric standard star obtained with the OSIRIS instrument at the Gran Telescopio Canarias (10.4m-GTC) at the Roque de los Muchachos Observatory (ORM, La Palma, Spain). The main characteristics of the observations are summarised in Table~\ref{tab:OSIRIS_data}. The closest seeing monitor to the GTC is the Differential Image Motion Monitor (DIMM) installed at the Telescopio Nazionale Galileo (3.58m-TNG). The local topography places the TNG $\sim$70\,m higher in altitude than the GTC, while both telescopes are separated by a straight-line distance of about 390\,m. Previous site-characterization studies at the ORM have shown that the seeing is generally homogeneous across the observatory, with differences between locations typically dominated by local effects rather than by large-scale atmospheric variations~[\citenum{1997Munoz-Tunon}]. Therefore, although the TNG DIMM provides the closest available seeing measurements to the GTC, local effects may still introduce differences between the atmospheric conditions sampled by the monitor and those affecting the GTC.

\begin{table*}[t]
\centering
\caption{Parameters of the OSIRIS long-slit data analysed in this study, reduced and provided by the GTC team.}
\label{tab:OSIRIS_data}
\vspace{1mm}
\tabcolsep 2.5pt
\begin{tabular}{ccccccccc}
\hline
Object & Obs. date & UTC & IT (s) & Grism & Slit width & Airmass & $\epsilon_{\mathrm{DIMM}}$  & $\epsilon_{\mathrm{OSIRIS}}$ \\
(1) & (2) & (3) & (4) & (5) & (6) & (7) & (8) & (9) \\
\hline
 Ross~640 & 04/10/2024 & 20:09:25 & 100 & R500B & 2.52 & 1.37 & 0.73$\pm$0.10 & 0.88  \\
 \hline
\end{tabular}
\vspace{1mm}

{\footnotesize\noindent\parbox{\linewidth}{
\raggedright
(1) Spectro-photometric standard star; (2) observing date (day/month/year); (3) observation time (UT); (4) integration time (s); (5) OSIRIS grism; (6) long-slit width (arcsec); (7) average air mass of the OSIRIS observation; (8) DIMM $\epsilon_{0}$ measurement, in arcsec, reported at the TNG monitor during the OSIRIS observations; (9) seeing (arcsec) along the OSIRIS line of sight, estimated from the DIMM values after correction for the airmass.
}}
\end{table*}

\begin{figure} [ht]
   \begin{center}
   \begin{tabular}{c} 
   \includegraphics[width=6.5cm]{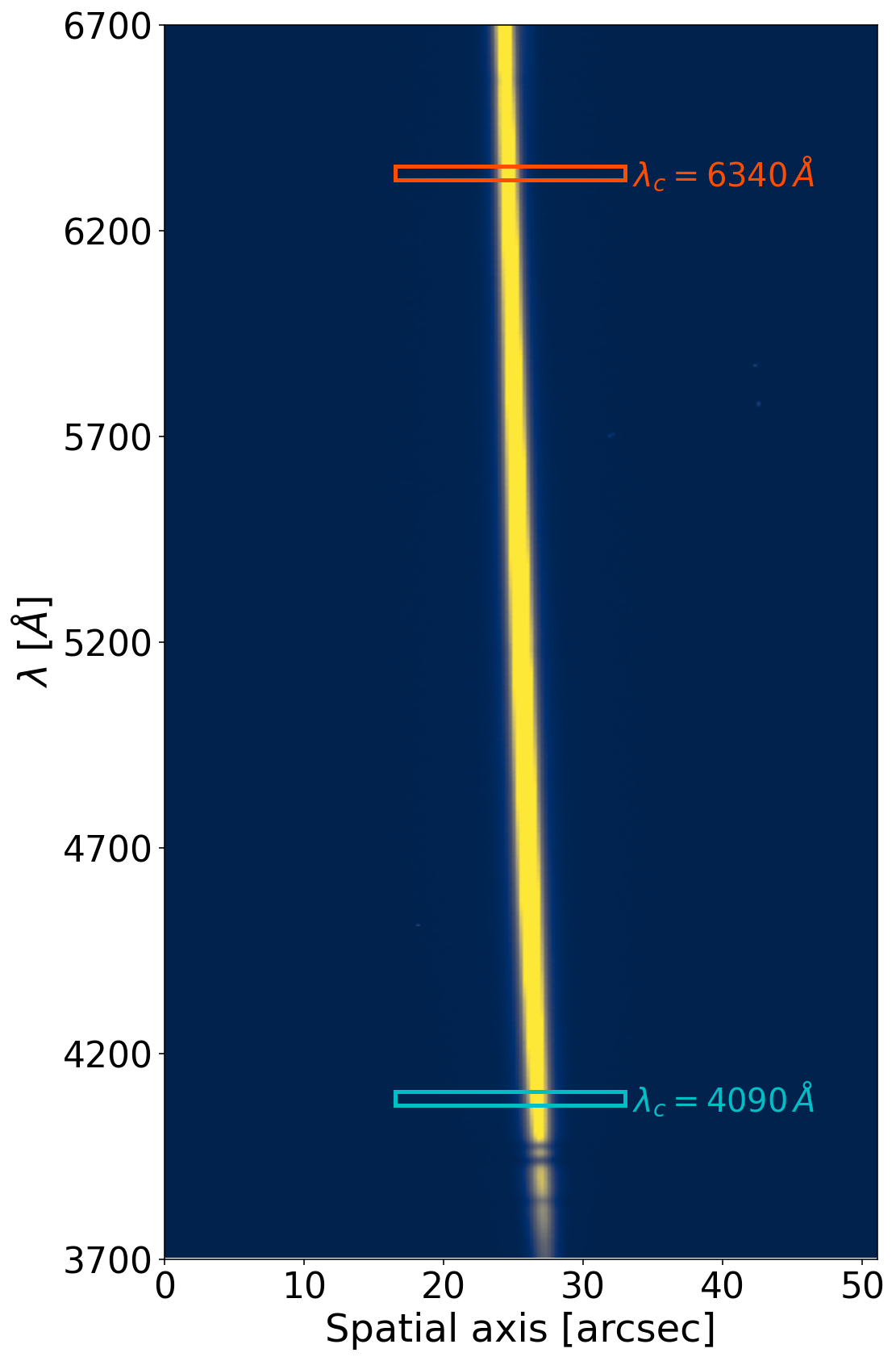}
   \includegraphics[width=9.75cm]{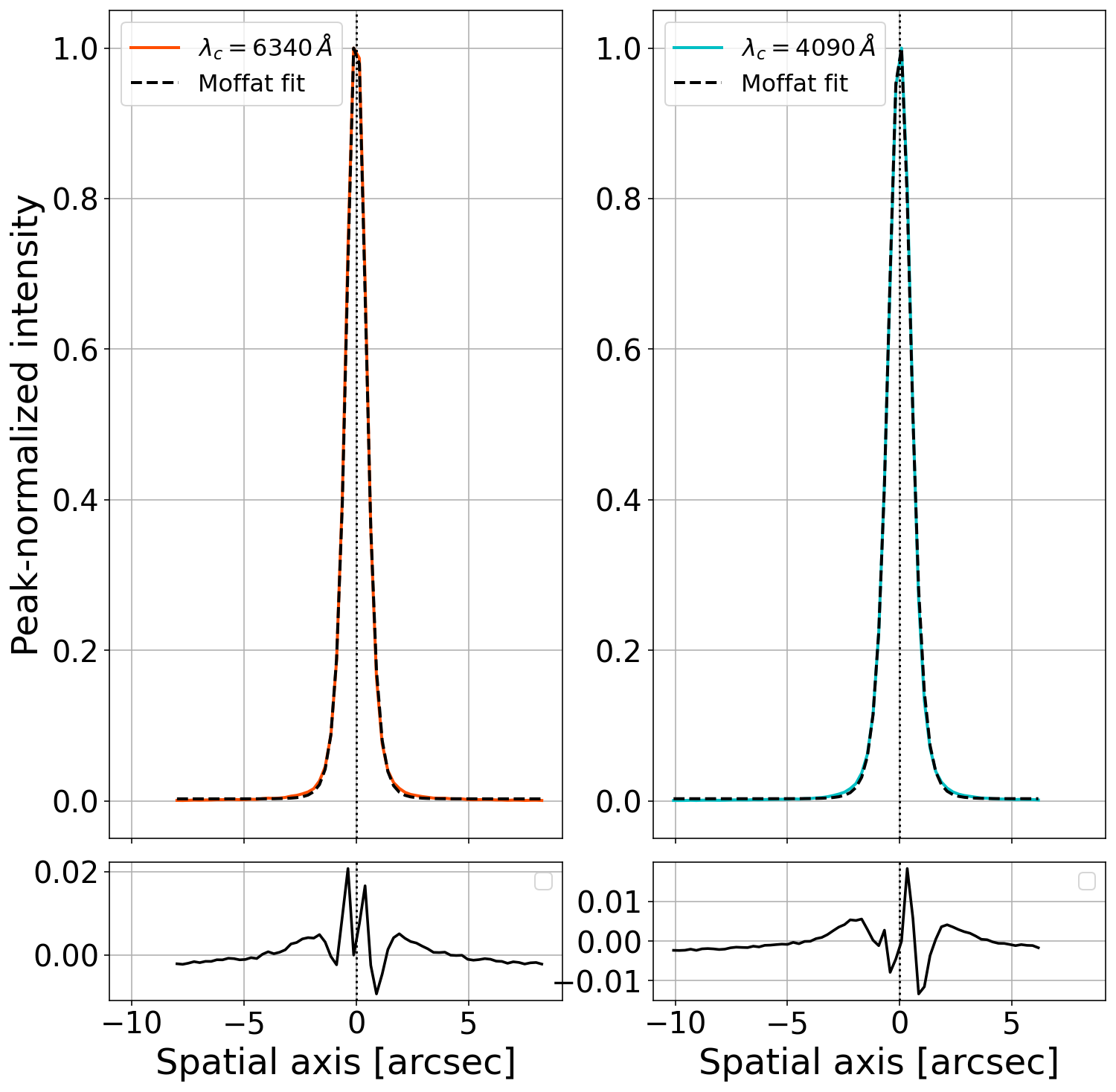}
   \end{tabular}
   \end{center}
   \caption[ ] 
   { \label{fig:OSIRIS_procedure} 
(Left) Example of a two-dimensional OSIRIS spectrum after sky subtraction. The vertical bright feature crossing the image corresponds to the observed stellar spectrum of Ross 640 spectro-photometric star. Colored rectangles indicate two of the spectral bands used to extract the spatial intensity profiles (PSFs) at different wavelengths. The labels indicate the central wavelength, $\lambda_c$, of each spectral band. (Right) Spatial intensity profiles extracted from the two highlighted spectral bands, normalized to the peak intensity. The dashed curves show the best-fit Moffat models, while the lower panels display the corresponding residuals.
} 
   \end{figure}

Figure \ref{fig:OSIRIS_procedure} shows typical long-slit data produced by the OSIRIS instrument for a single point source, processed up to the sky-subtraction stage. The starlight is dispersed along the spectral direction (the Y-axis in Fig. \ref{fig:OSIRIS_procedure}), while the spatial direction is perpendicular to the dispersion axis (the X-axis in Fig. \ref{fig:OSIRIS_procedure}). 

The 2D spectra were divided into consecutive sections along the spectral direction, each spanning approximately 10 pixels, corresponding to a wavelength interval of about 16\,\AA, to simulate narrow-band filters. For each section, the median spatial profile was computed to obtain a one-dimensional PSF profile at the corresponding central wavelength. These PSF profiles were then modeled using a Moffat function implemented through the \texttt{astropy.modeling} package. The FWHM derived from the Moffat model was adopted as a measure of the OSIRIS IQ at each wavelength. Figure~\ref{fig:OSIRIS_procedure} illustrates the procedure. The left panel shows the sky-subtracted two-dimensional spectrum, together with two representative spectral bands used to extract the spatial profiles. The right panels display the corresponding spatial intensity profiles at two different wavelengths, along with their best-fitting Moffat models and residuals. Although the residuals indicate the presence of additional low-level structures not captured by the Moffat model, these deviations are mainly confined to the profile wings and are not expected to significantly affect the determination of the FWHM. A more detailed characterization of the spatial profiles would be required to assess the potential impact of slit effects and other instrumental contributions. Nevertheless, as a first-order approximation, we perform a preliminary analysis assuming that the observed IQ is primarily dominated by turbulence-induced image degradation, while instrumental aberrations and slit-related effects remain secondary.

\begin{figure} [ht]
   \begin{center}
   \begin{tabular}{c} 
   \includegraphics[width=14cm]{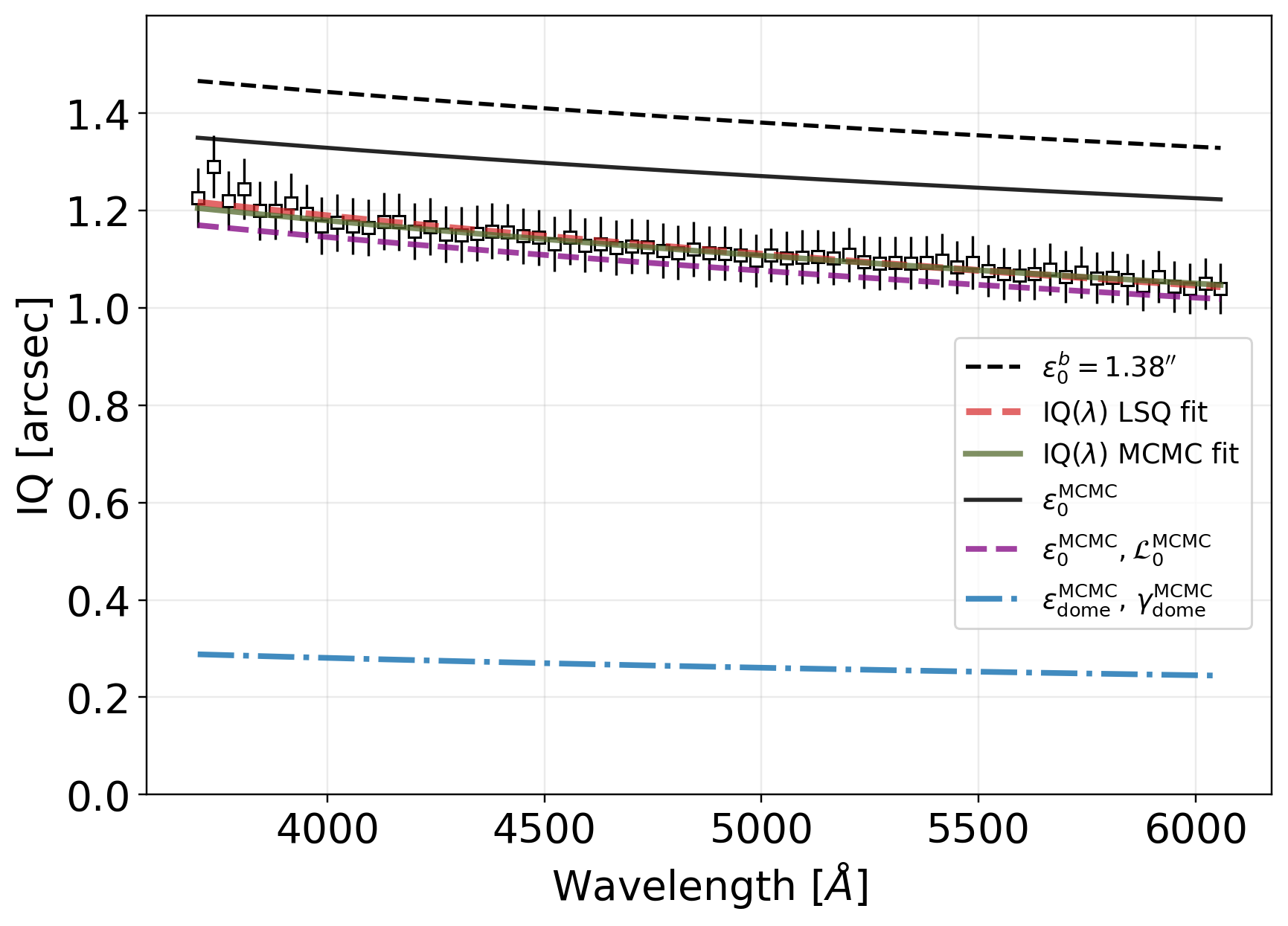} 
   \end{tabular}
   \end{center}
   \caption[ ] 
   { \label{fig:iq_lambda} 
Wavelength dependence of the IQ (open black squares) at the OSIRIS focal plane, derived from the long-slit observations of a bright spectrophotometric standard star. The error bars correspond to the uncertainties in the IQ measurements derived from Monte Carlo simulations assuming Poisson noise in the PSF profiles, with the errors propagated from the fitted Moffat parameters. The black dashed line shows the $\lambda^{-1/5}$ seeing dependence, scaled from the best $\epsilon_0^{b}$ obtained from the LSQ fit initialized with the DIMM seeing projected along the OSIRIS line of sight ($\epsilon_{\rm OSIRIS}$; Table~\ref{tab:OSIRIS_data}). The red dashed line represents the best-fitting least-squares (LSQ) model following Eq.~\ref{eq:Tokovinin}, characterized by the parameters $\epsilon_{0}^{b}$ and \LO$^{b}$ listed in Table~\ref{tab:derivedparameters}. The green solid line corresponds to the MCMC best-fitting model, which includes both atmospheric and dome-seeing contributions to the IQ. The black solid line shows the seeing curve obtained from the MCMC analysis, while the purple and blue curves represent the atmospheric and dome-seeing contributions to the IQ derived from the same analysis. The corresponding MCMC-derived values for the seeing, \LO, dome seeing, and $\gamma$ are also listed in Table~\ref{tab:derivedparameters}.
} 
\end{figure}

Figure \ref{fig:iq_lambda} shows the IQ measurements derived from the different narrow-band intensity profiles for the long-slit OSIRIS observation (see Table \ref{tab:OSIRIS_data}). Despite the reduced spatial information compared with IFS observations, the measured wavelength behavior follows the expected turbulence-driven trends (i.e., Eq. \ref{eq:Tokovinin}), supporting the feasibility of extending focal-plane turbulence diagnostics to long-slit spectroscopic data.

   \begin{figure} [ht]
   \begin{center}
   \begin{tabular}{c} 
   \includegraphics[width=14cm]{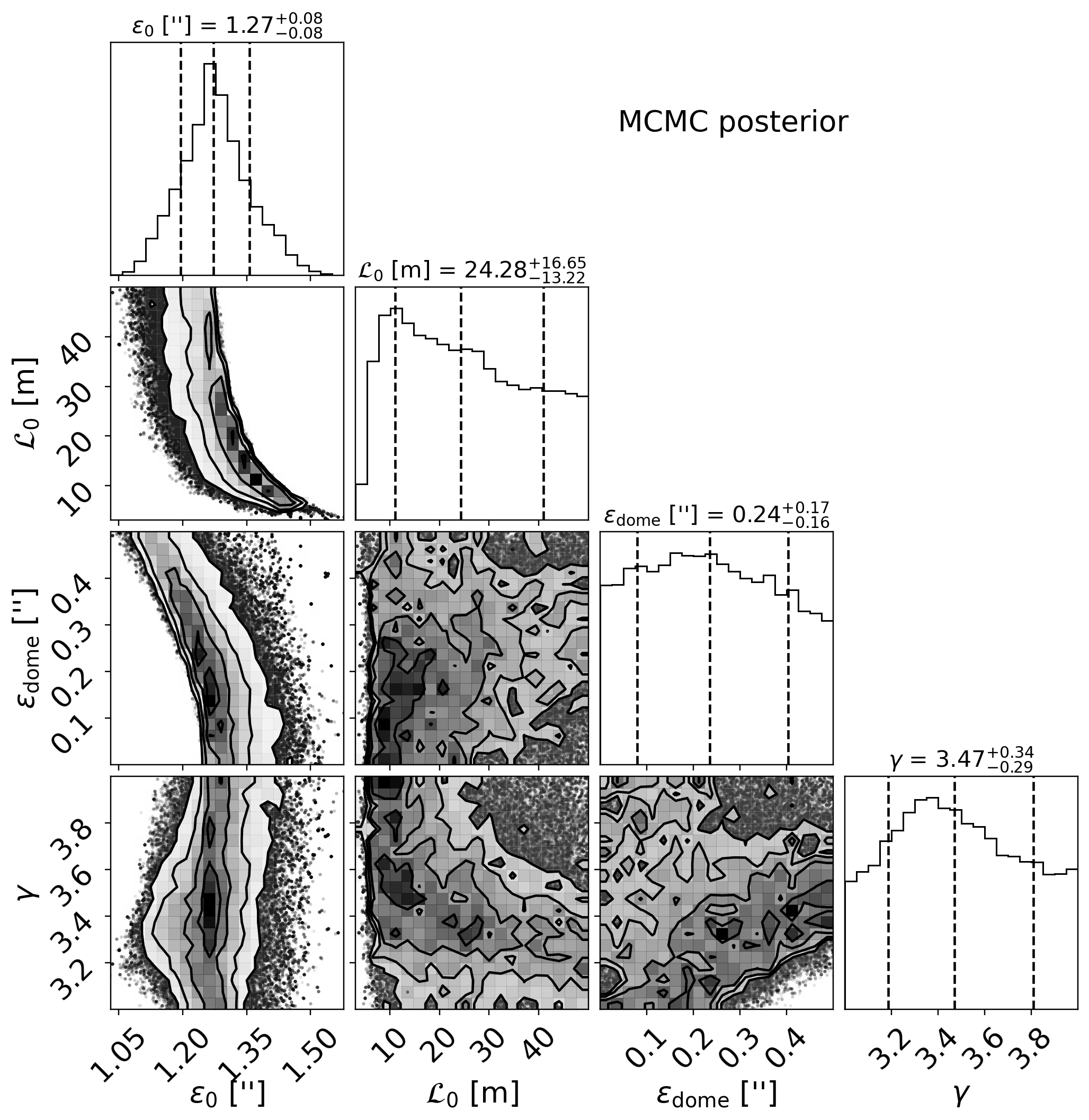}
   \end{tabular}
   \end{center}
   \caption[ ] 
   { \label{fig:MCMCPosterior} 
The posterior probability distributions of the free parameters obtained from the MCMC analysis. The diagonal panels show the marginalized one-dimensional posterior distributions, while the off-diagonal panels display the two-dimensional projections of the posterior distributions, illustrating the correlations and degeneracies between parameters. The values quoted above each distribution correspond to the median of the posterior distribution, with uncertainties derived from the 16th and 84th percentiles ($\approx 1\sigma$). See also Table~\ref{tab:derivedparameters}.
} 
   \end{figure} 

\begin{table*}
\caption{Atmospheric parameters derived from the measured IQ at different wavelengths from OSIRIS@GTC long-slit observations of spectrophotometric standard stars.}
\label{tab:derivedparameters}
\begin{center}
\vspace{1mm}
\tabcolsep 8.5pt
\begin{tabular}{c|cc|cccc|}
 & \multicolumn{2}{c|}{\bf Negligible} & \multicolumn{4}{c|}{\bf Considering} \\ [3pt] 
 & \multicolumn{2}{c|}{\bf dome-seeing} & \multicolumn{4}{c|}{\bf dome-seeing} \\ [3pt]
\hline \rule{0pt}{12pt}
{\bf Standard star}   
& $\epsilon_{0}^{b}$ 
& $\mathcal{L}_{0}^{b}$  
& $\epsilon_{0}^{{\mathrm{MCMC}}}$ 
& $\mathcal{L}_{0}^{{\mathrm{MCMC}}}$ 
& $\epsilon_{\rm dome}^{{\mathrm{MCMC}}}$ 
& $\gamma^{{\mathrm{MCMC}}}$ \\ [3pt]

(1) & (2) & (3) & (4) & (5) & (6) & (7) \\ [3pt]
\hline \rule{0pt}{12pt}
Ross640  & 1.38$\pm$0.09 & 12.4$\pm$9.5  & 1.27$_{-0.08}^{+0.08}$ & 24.3$_{-13.2}^{+16.6}$ & 0.23$_{-0.15}^{+0.17}$ & 3.47$_{-0.28}^{+0.34}$ \\ [3pt]
\hline 
\end{tabular}
\end{center}
\medskip
\footnotesize
Columns correspond to:  
(1) Observed OSIRIS spectrophotometric star; 
(2) and (3) best-fit values of $\epsilon_{0}$ (arcsec) and \LO\ (m), obtained by fitting the predicted $\epsilon_{LE}$ (Eq.~\ref{eq:Tokovinin}) to the MUSE-measured IQ under the assumption of negligible dome seeing; 
(4)--(7) best-fit values of $\epsilon_{0}$, \LO\ (m), $\epsilon_{\rm dome}$ (arcsec), and the power-law index $\gamma$ (Eq.~\ref{eq:dome_lambda}), including the dome-seeing contribution.
\end{table*}

We first considered the simplified case in which the observed IQ is entirely driven by atmospheric turbulence with a finite outer scale. The empirical $\mathrm{IQ}_{\rm OSIRIS}$ measurements were fitted using the analytical model of~\citenum{2002Tokovinin} and a weighted non-linear least-squares (LSQ) method, taking into account the uncertainties of each data point through a $\chi^2$ minimization. The fit was initialized using the TNG DIMM seeing projected along the OSIRIS line of sight as the initial value for the seeing parameter, and a typical \LO\ value of 15 m for the ORM~[\citenum{1999Wilson}] as the initial outer-scale estimate. Figure~\ref{fig:iq_lambda} shows the resulting best-fitting model (purple curve), characterized by the parameters $\epsilon_0^{b}$ and \LO$^{b}$ (Table~\ref{tab:derivedparameters}). Both derived atmospheric turbulence parameters are consistent with the range of seeing and outer-scale values previously reported for the ORM (e.g. [\citenum{1999Wilson}], [\citenum{1997Munoz-Tunon}]).

Although the design and operation of large telescopes aim to minimize turbulence generated within their enclosures, residual turbulence within the dome is generally unavoidable, and the resulting dome seeing contributes to the final IQ at the focal plane. For the GTC enclosure, no specific dome-seeing model or empirical measurements are currently available, at least to our knowledge. We therefore adopt the approach originally introduced for the CFHT~[\citenum{1991Racine}] and more recently applied to the VLT by~[\citenum{2024Garcia-Lorenzo}], in which the total IQ is described as the quadratic sum of the atmospheric and dome-seeing contributions (Eqs.~\ref{eq:IQ_total} and ~\ref{eq:dome_lambda}). 
Since including the dome-seeing component introduces additional degrees of freedom and potential parameter degeneracies, we complement the LSQ analysis with a Bayesian Markov Chain Monte Carlo (MCMC) approach to explore the posterior distributions of the model parameters. In the absence of empirical dome seeing measurements at GTC, we adopt an initial dome-seeing contribution of 0.25 arcsec, corresponding to the typical value derived for the VLT using the same methodology. The dome-seeing amplitude is allowed to vary between 0 and 0.5 arcsec, while the atmospheric parameters are constrained around the LSQ best-fit values. This setup tries to reduce parameter degeneracies and guides the exploration toward physically plausible solutions. The MCMC analysis provides posterior distributions for $\epsilon_0$, \LO , $\epsilon_{\rm dome}$, and $\gamma_{\mathrm{dome}}$ (Fig. ~\ref{fig:MCMCPosterior}), from which we derive parameter estimates, uncertainties, and correlations. We adopt the median of each posterior distribution as the reference value (see Table~\ref{tab:derivedparameters}), while the 16th and 84th percentiles define the central 68\% credible interval (equivalent to the 1$\sigma$ interval for a Gaussian distribution).

The atmospheric parameters derived from this analysis are compatible with the typical turbulence conditions reported for the ORM, and the inferred dome seeing values appear realistic when compared with those estimated for other telescopes. Nevertheless, the absence of simultaneous independent measurements prevents a rigorous validation of the retrieved parameters. The present study should therefore be considered a preliminary demonstration of the potential applicability of the methodology proposed by [\citenum{2024Garcia-Lorenzo}] to long-slit spectroscopic observations.

\subsection{Multi-band imaging instruments}

Simultaneous multi-band imagers have become an important class of imaging instruments, particularly for time-domain and exoplanet science. They provide an observational framework that naturally satisfies the requirements of the turbulence-diagnostic methodology described above. By splitting the incoming beam into multiple optical channels and simultaneously recording images in several broad-band filters, these instruments deliver measurements obtained under essentially identical atmospheric and instrumental conditions. Instruments such as MuSCAT\footnote{https://research.iac.es/OOCC/iac-managed-telescopes/telescopio-carlos-sanchez/muscat2/}, HiPERCAM\footnote{https://www.gtc.iac.es/instruments/hipercam/hipercam.php}, and GROND\footnote{https://www.mpe.mpg.de/~jcg/GROND/} have demonstrated the feasibility of obtaining high-precision multi-band observations on telescopes with apertures ranging from $\sim$1.5 to 10.4 m. In this context, the wavelength dependence of the IQ, or equivalently of the PSF, measured from the different channels could, in principle, provide constraints on atmospheric and dome-induced turbulence parameters analogous to those obtained from IFS, albeit with a more limited spectral sampling.

To explore the applicability of the method to simultaneous multi-band imagers, we take advantage of the continuous spectral coverage provided by MUSE data cubes. Rather than analysing observations obtained with a dedicated multi-channel instrument, we reconstruct a small number of synthetic broad-band images from the IFS data. This approach allows us to emulate the observational conditions of simultaneous multi-band imagers while retaining the homogeneous atmospheric and instrumental conditions inherent to the original MUSE observations. Since the same MUSE observations can be analysed both with the original narrow-band methodology [\citenum{2024Garcia-Lorenzo}] and with a reduced set of synthetic broad-band filters, the retrieved turbulence parameters can be compared directly. This provides a controlled test of whether a limited number of simultaneous broad-band measurements can recover results consistent with those obtained from the full IFS spectral information.

\begin{figure} [ht]
   \begin{center}
   \begin{tabular}{c} 
   \includegraphics[width=5.75cm]{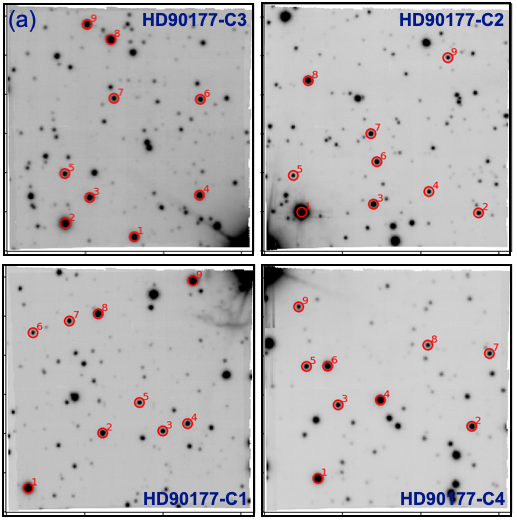} \hspace{0.5cm}
   \includegraphics[width=9.75cm]{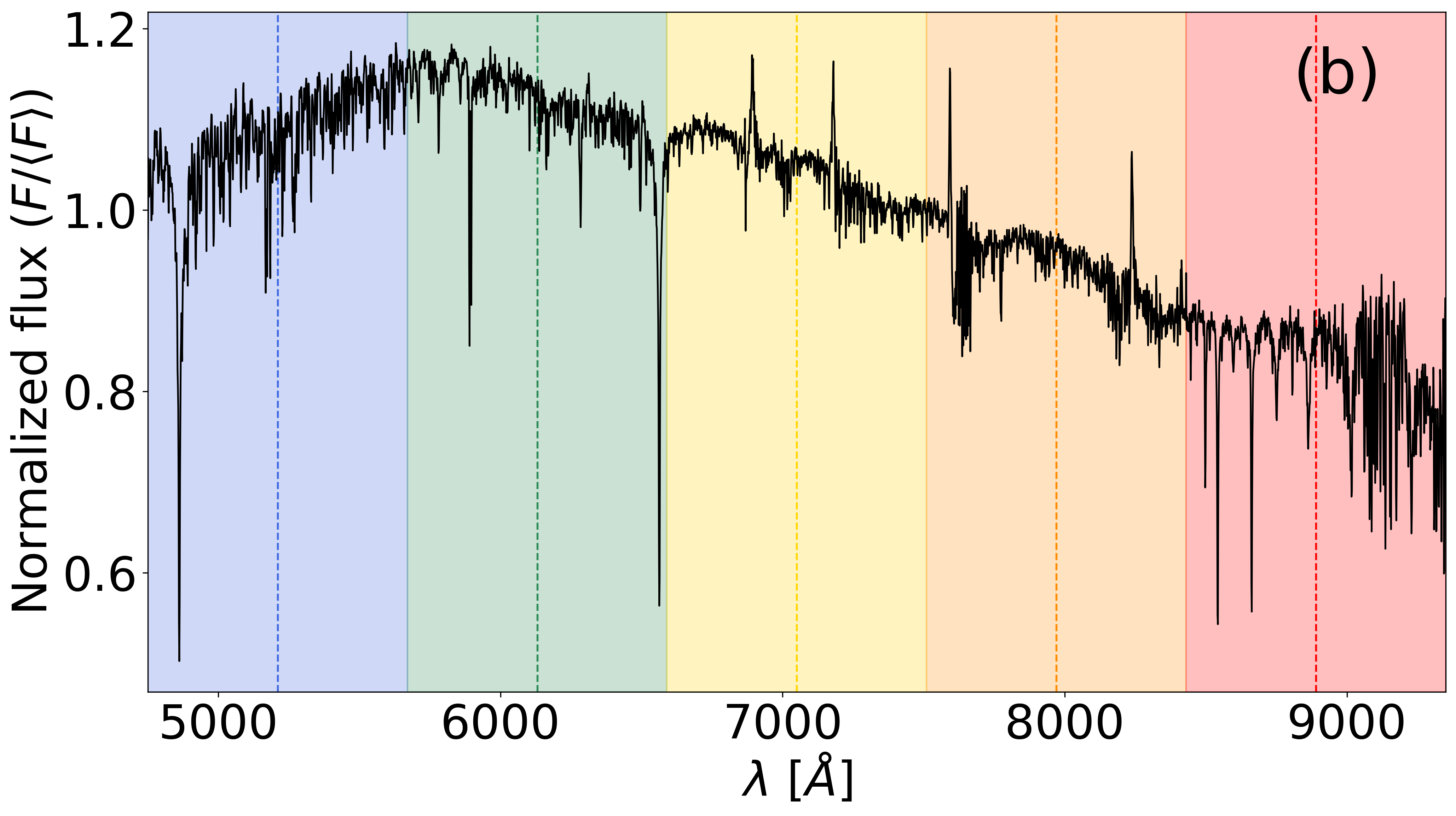}
   \end{tabular}
   \end{center}
   \caption[ ] 
   { \label{fig:multi-image} 
(a) Combined field of view reconstructed from the four MUSE data cubes analyzed in this work. The brightest star, HD~90177, is located near the center of the mosaic. The stars selected for further analysis in each field are marked with red circles and labels. (b) Example of a stellar spectrum for star 2 in the HD90177-C4 field, normalized to its mean flux and extracted within a circular aperture of 1.2 arcsec radius. The shaded regions indicate the five synthetic broadband filters adopted to emulate the spectral sampling of a simultaneous multi-band imaging instrument. The vertical dashed lines mark the central wavelength of each filter.
} 
   \end{figure}

\begin{figure} [ht]
   \begin{center}
   \begin{tabular}{c} 
   \includegraphics[width=8.25cm]{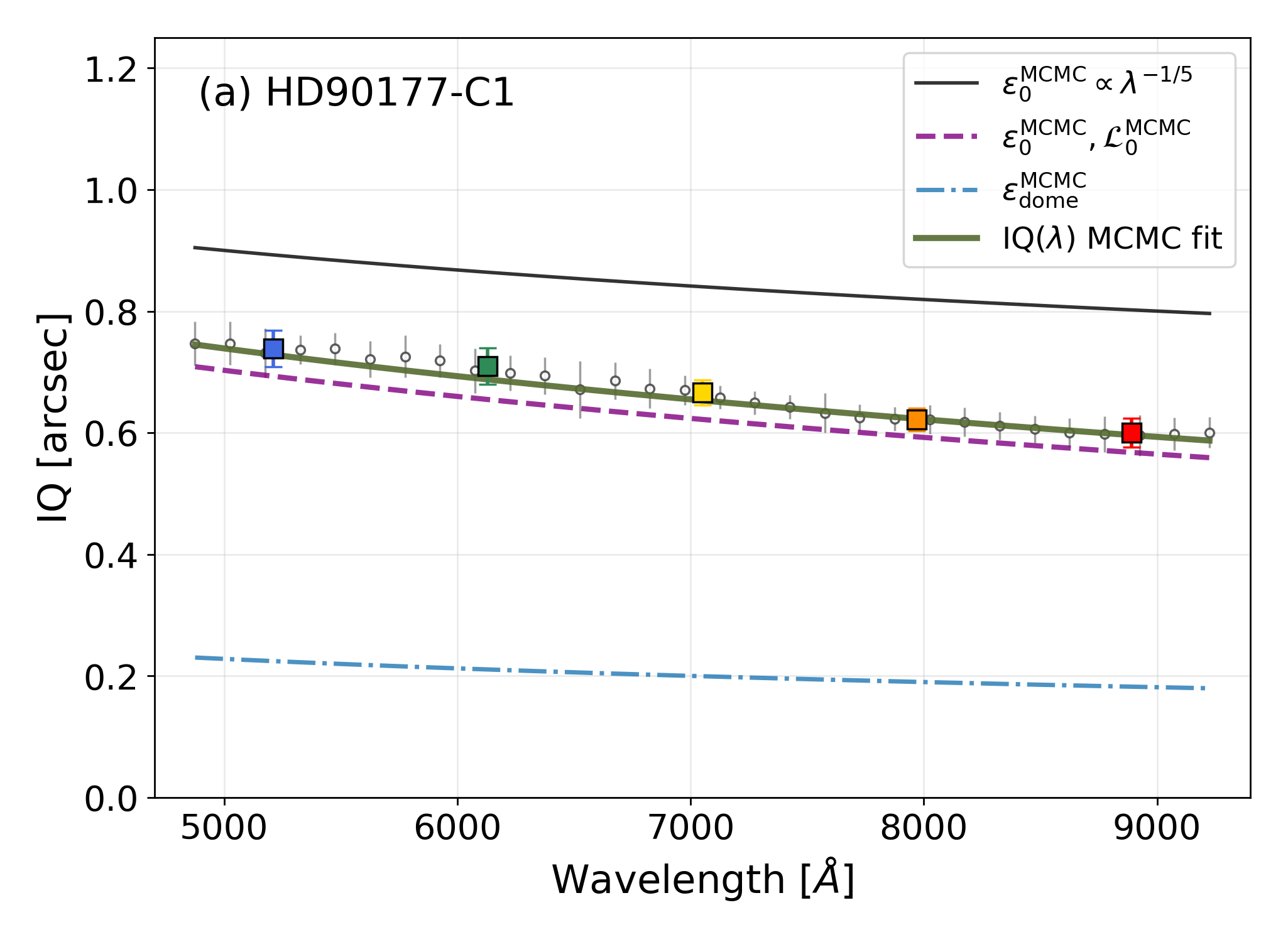}
   \includegraphics[width=8.25cm]{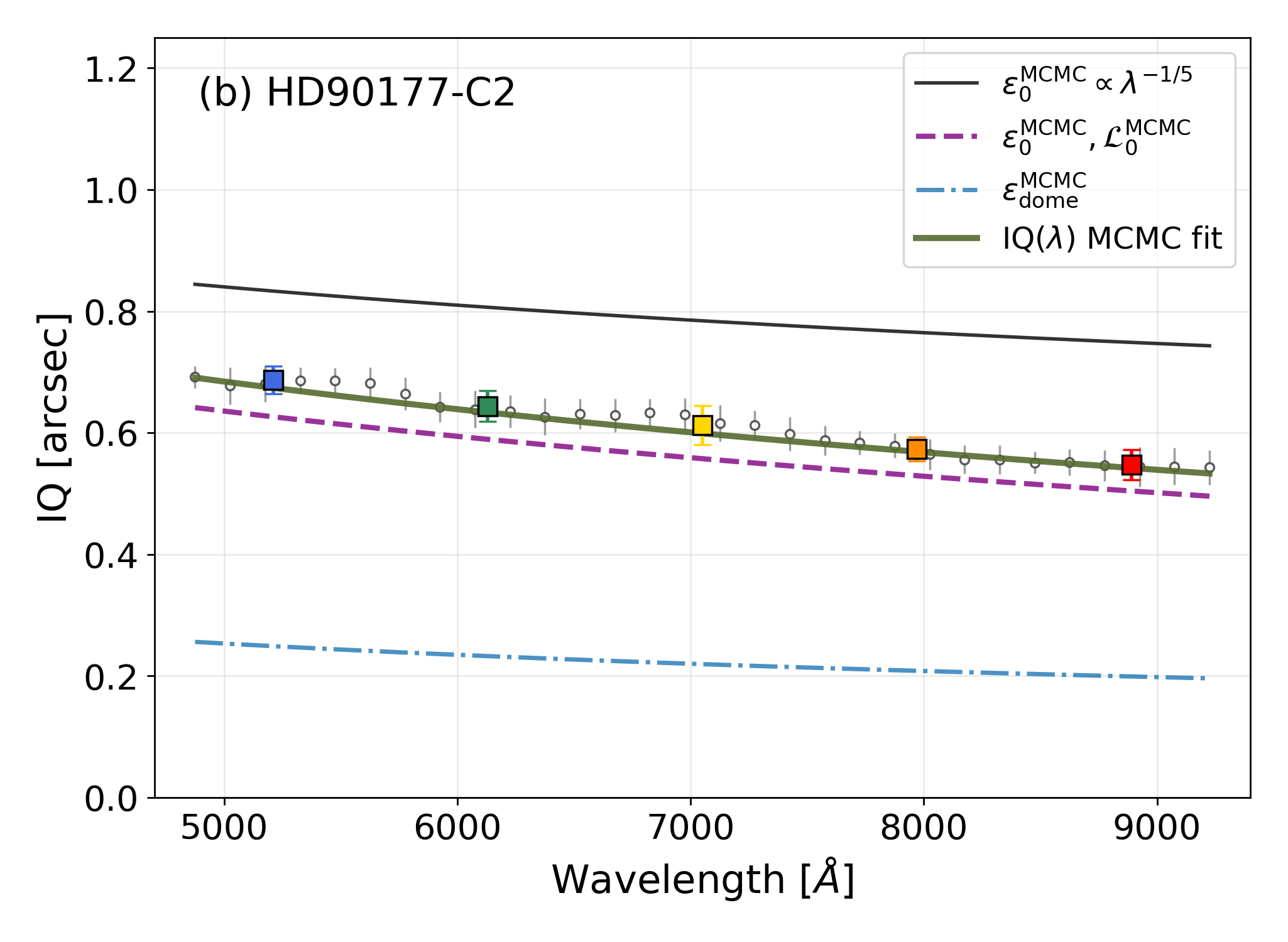} \\
   \includegraphics[width=8.25cm]{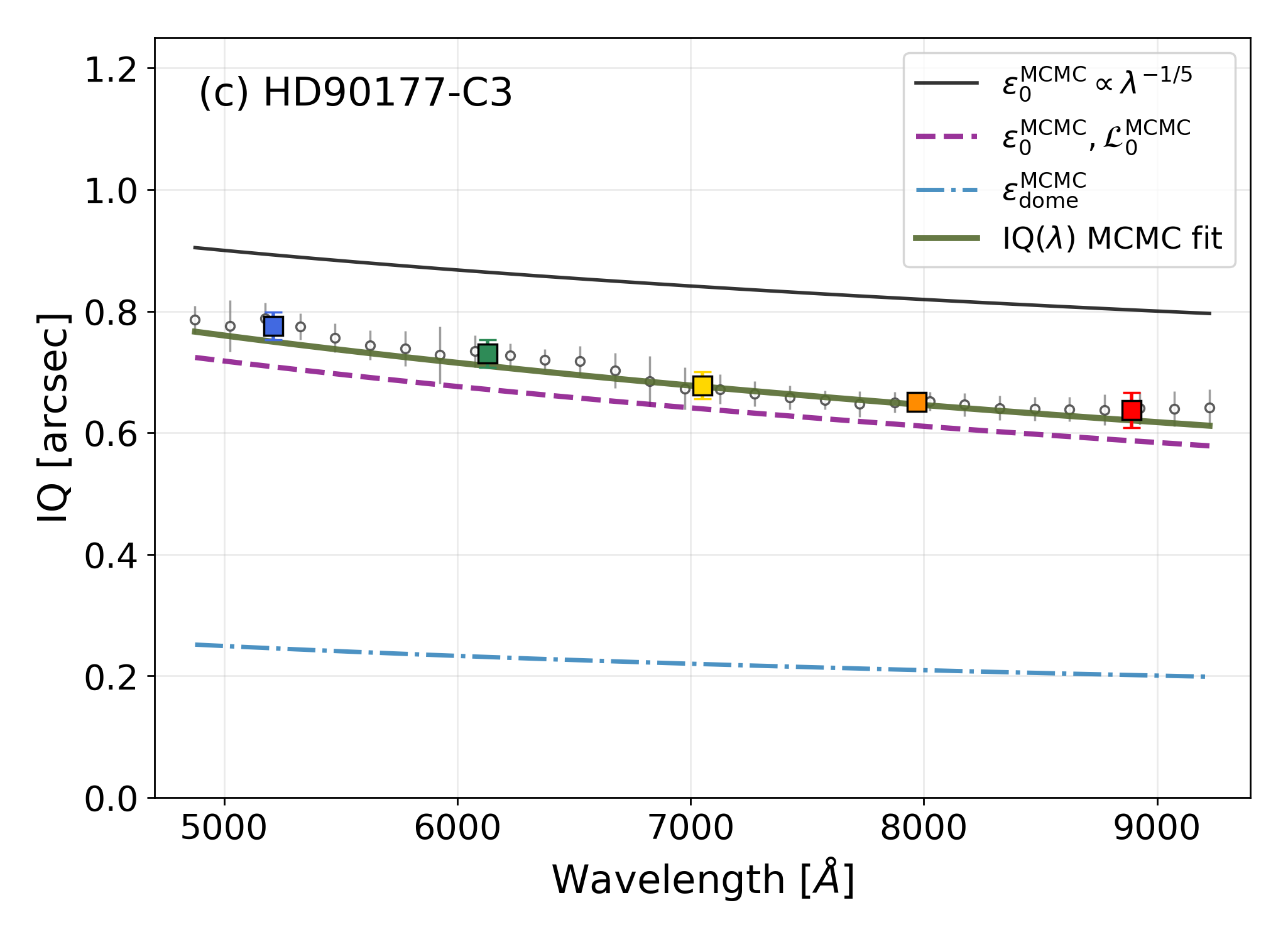}
    \includegraphics[width=8.25cm]{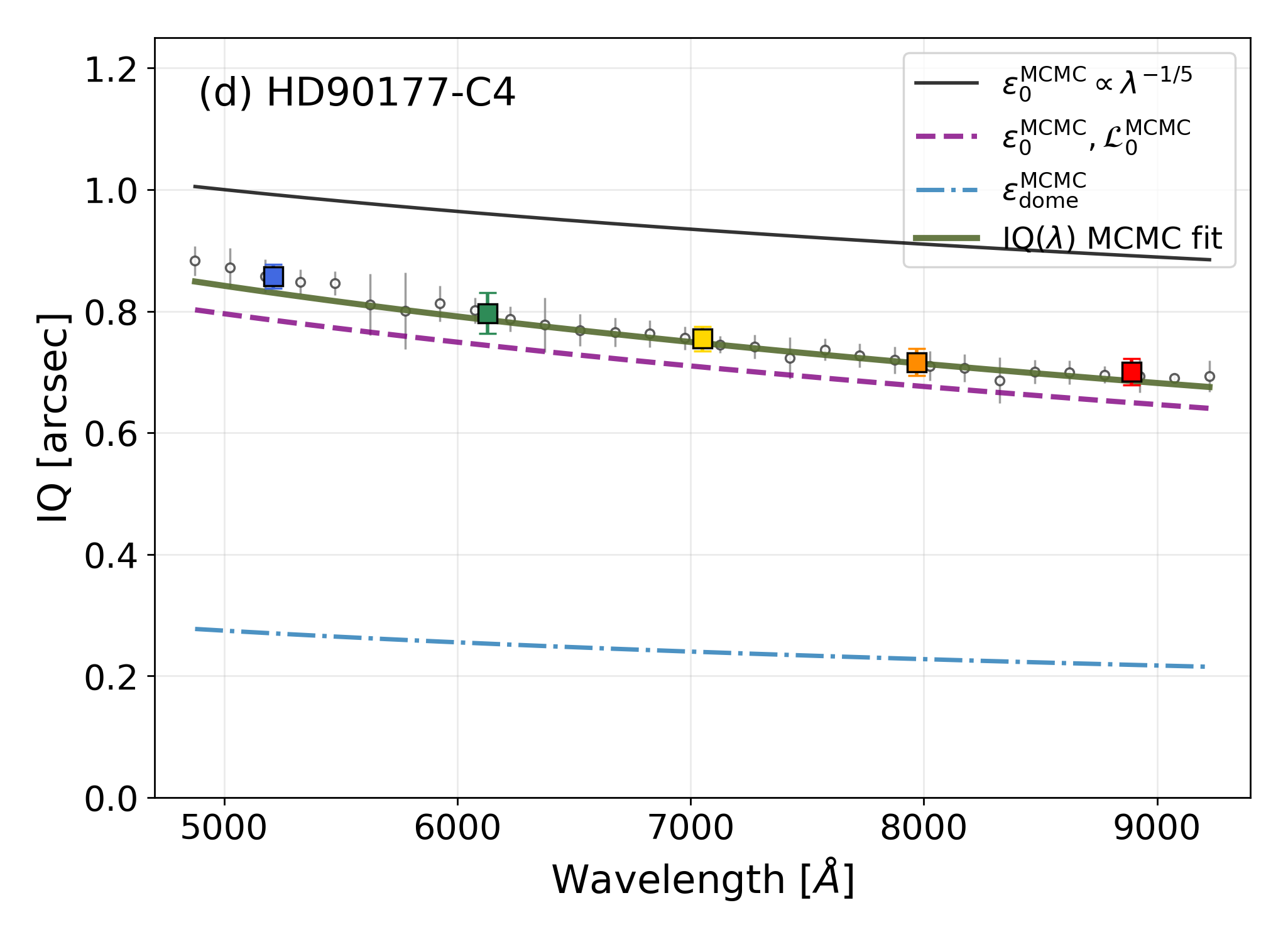}
   \end{tabular}
   \end{center}
   \caption[ ] 
   { \label{fig:IQ_5filters} 
Wavelength dependence of the IQ at the MUSE focal plane for the four MUSE data cubes analyzed in this work, as indicated in each panel. The black open squares with error bars show the mean IQ($\lambda$) measurements obtained from 30 narrow-band filters across the MUSE spectral range (see [\citenum{2026Garcia-Lorenzo}], these proceedings). The colored symbols represent the IQ measurements obtained from the five synthetic broad-band filters shown in Fig.~\ref{fig:multi-image}(b), designed to emulate the spectral sampling of a simultaneous multi-band imaging instrument. The black solid line shows the seeing curve derived from the MCMC fit to the five broadband IQ measurements, while the dashed purple and dot-dashed blue curves represent the corresponding atmospheric and dome-seeing contributions, respectively. The green solid line corresponds to the best-fitting MCMC model, including both atmospheric and dome-seeing contributions to the IQ. The corresponding MCMC-derived values of the seeing, \LO, dome-seeing contribution, and $\gamma$ are listed in Table~\ref{tab:mcmc_result}.} 
   \end{figure}

As a representative example, we consider the four MUSE data cubes previously studied by García-Lorenzo et al. ([\citenum{2026Garcia-Lorenzo}], these proceedings), where the wavelength dependence of the IQ was characterized using 30 narrow-band filters spanning the MUSE spectral range. Figure~\ref{fig:multi-image}(a) shows the reconstructed field obtained by combining the four individual MUSE pointings, thereby recovering the full observed region around HD~90177 and placing this bright star near the center of the mosaic. In the present work, we adopt a configuration inspired by instruments such as HiPERCAM and define five synthetic broad-band filters of equal width within the MUSE wavelength coverage. Specifically, the MUSE spectral range ($\sim4750$--$9350$ \AA) is divided into five contiguous broad-band filters, each 920 \AA\ wide, with the first filter centered at 5210 \AA\ (Fig.~\ref{fig:multi-image}(b)). Although these filters neither reproduce the exact HiPERCAM passbands nor fully overlap with the HiPERCAM filter set, they provide a realistic approximation of the spectral sampling achievable with current simultaneous multi-band imagers. Following the procedure described in [\citenum{2026Garcia-Lorenzo}], nine stars were selected in each MUSE data cube for the analysis. For each star and filter, the PSF was modeled with a Moffat profile. The resulting error-weighted mean IQ, computed from the nine stellar measurements, was adopted as the representative IQ($\lambda$) relation for each field (Fig.~\ref{fig:IQ_5filters}), while the corresponding weighted scatter among the stellar measurements at each wavelength was taken as the uncertainty.

The resulting IQs($\lambda$) were fitted following the methodology described by [\citenum{2026Garcia-Lorenzo}]. Briefly, the observed IQ($\lambda$) behavior was modeled as the combined contribution of atmospheric turbulence (Eq.~\ref{eq:Tokovinin}) and dome-induced turbulence (Eq.~\ref{eq:dome_lambda}), whose quadratic combination is given by Eq.~\ref{eq:IQ_total}. As in [\citenum{2026Garcia-Lorenzo}], a weighted non-linear LSQ fit was first performed to obtain an initial estimate of the model parameters. These values were subsequently used as starting points for a MCMC analysis, whose posterior distributions were used to derive the best-fitting parameters and their associated uncertainties. The resulting MCMC estimates for each of the four MUSE data cubes analyzed in this work are summarized in Table~\ref{tab:mcmc_result}.

\begin{table*}
\centering
\caption{Atmospheric parameters derived from the measured IQ at different wavelengths from MUSE data cubes.}
\label{tab:mcmc_result}
\vspace{8pt}

\begin{tabular}{c|cc|cccc|}
 & \multicolumn{2}{c|}{\bf Negligible} & \multicolumn{4}{c|}{\bf Considering} \\ [3pt] 
 & \multicolumn{2}{c|}{\bf dome-seeing} & \multicolumn{4}{c|}{\bf dome-seeing} \\ [3pt]
\hline \rule{0pt}{12pt}
{\bf MUSE Field}  
& $\epsilon_{0}^{b}$ 
& $\mathcal{L}_{0}^{b}$  
& $\epsilon_{0}^{{\mathrm{MCMC}}}$ 
& $\mathcal{L}_{0}^{{\mathrm{MCMC}}}$ 
& $\epsilon_{\rm dome}^{{\mathrm{MCMC}}}$ 
& $\gamma$ \\ [3pt]

(1) & (2) & (3) & (4) & (5) & (6) & (7) \\ [3pt]
\hline \rule{0pt}{12pt}
HD90177-C1 & 0.98$\pm$0.05 & 10.6$\pm$3.6  & 0.90$_{-0.08}^{+0.07}$ & 14.08$_{-5.2}^{+8.6}$ & 0.20$_{-0.13}^{+0.14}$ & 3.44$_{-0.26}^{+0.34}$ \\ [3pt]
HD90177-C2 & 0.94$\pm$0.04 & 9.1$\pm$2.5  & 0.84$_{-0.11}^{+0.07}$ & 11.7$_{-4.3}^{+8.9}$ & 0.22$_{-0.14}^{+0.14}$ & 3.41$_{-0.25}^{+0.34}$ \\ [3pt]
HD90177-C3 & 0.98$\pm$0.05 & 12.8$\pm$5.2  & 0.90$_{-0.09}^{+0.07}$ & 17.2$_{-6.9}^{+9.4}$ & 0.22$_{-0.15}^{+0.16}$ & 3.46$_{-0.26}^{+0.31}$ \\ [3pt]
HD90177-C4 & 1.10$\pm$0.05 & 10.7$\pm$4.1  & 1.00$_{-0.09}^{+0.08}$ & 15.1$_{-6.1}^{+9.2}$ & 0.24$_{-0.15}^{+0.16}$ & 3.43$_{-0.25}^{+0.33}$ \\ [3pt]
\hline
\end{tabular}
\vspace{8pt}

\tablefoot{Columns correspond to: 
(1) observed MUSE field; 
(2) and (3) best-fit values of $\epsilon_{0}$ (arcsec) and \LO\ (m), obtained by fitting the predicted $\epsilon_{LE}$ (Eq.~\ref{eq:Tokovinin}) to the MUSE-measured IQ under the assumption of negligible dome seeing; 
(4)--(7) best-fit values of $\epsilon_{0}$, \LO\ (m), $\epsilon_{\rm dome}$ (arcsec), and the power-law index $\gamma$ (Eq.~\ref{eq:dome_lambda}), including the dome-seeing contribution.
}
\end{table*}

   \begin{figure} [ht]
   \begin{center}
   \begin{tabular}{c} 
   \includegraphics[width=14cm]{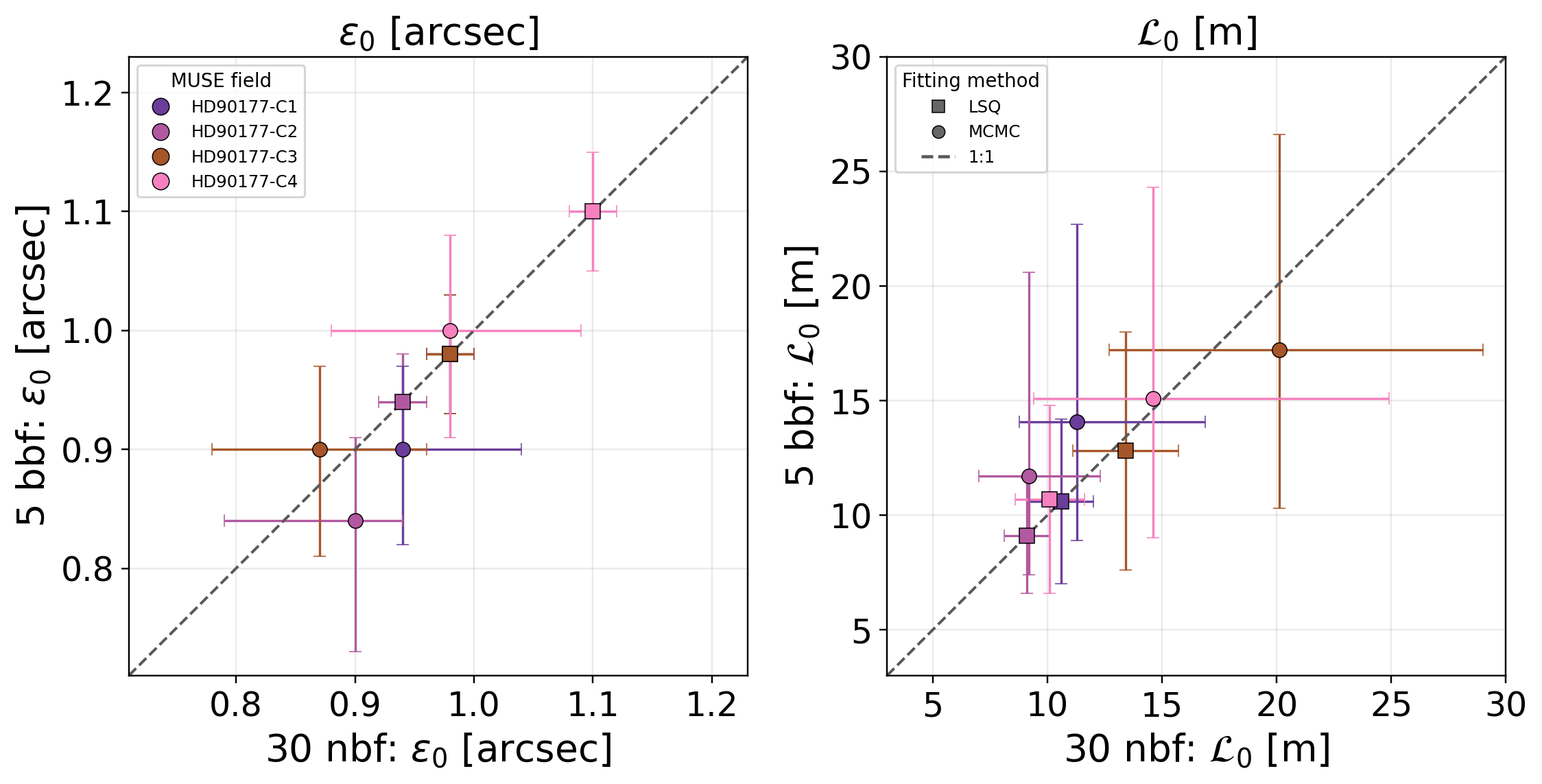} \\
   \includegraphics[width=14cm]{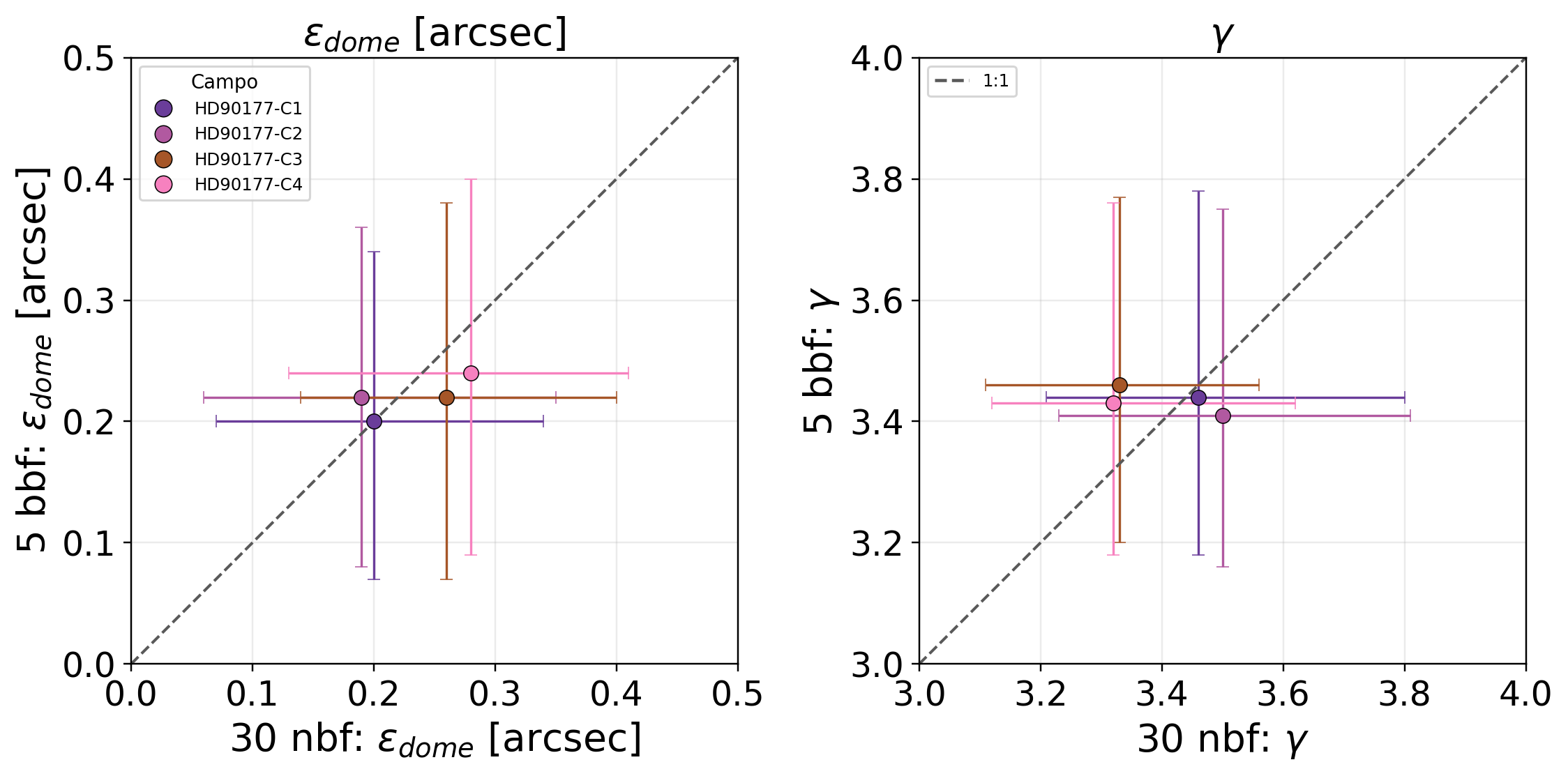} 
   \end{tabular}
   \end{center}
   \caption[ ] 
   { \label{fig:par_compara_5_30} 
   Comparison of turbulence parameters derived from fitting the IQ wavelength dependence over the MUSE spectral range using either 30 narrow-band filters or 5 synthetic broad-band filters. The upper panels show the atmospheric parameters ($\epsilon_0$ and \LO), while the lower panels show the dome-turbulence parameters ($\epsilon_{\mathrm{dome}}$ and $\gamma$). Squares represent LSQ fits assuming purely atmospheric image degradation, whereas circles correspond to MCMC estimates including both atmospheric and dome-induced turbulence contributions. Error bars indicate the associated parameter uncertainties, and the dashed line marks the one-to-one relation. The close agreement between both approaches demonstrates that five simultaneous broad-band measurements provide turbulence constraints comparable to those obtained from the full 30-filter spectral sampling. Colors indicate the corresponding MUSE field, as shown in the legend.} 
   \end{figure} 

To assess the impact of the reduced spectral sampling, we compare the turbulence parameters derived from the fits to these five broadband IQ measurements with those obtained by [\citenum{2026Garcia-Lorenzo}] from the same MUSE data cubes using 30 narrow-band filters sampling the same spectral range. The comparison (see Fig. \ref{fig:par_compara_5_30}) indicates that the parameters derived from both approaches are consistent within the current uncertainties. These uncertainties are partly driven by the fact that the analyzed observations were not specifically acquired for turbulence characterization and by the absence of a dome-seeing model specifically calibrated for the VLT enclosure where the MUSE data were obtained, which introduces additional uncertainty in the separation of atmospheric and dome-induced turbulence contributions. The adopted dome-seeing model should therefore be regarded as a phenomenological description rather than a physically validated representation of turbulence inside the VLT enclosure.

Nevertheless, the comparison suggests that five simultaneous broad-band measurements are sufficient to recover turbulence parameters consistent with those obtained from the full IFS spectral sampling. This result supports the feasibility of extending the methodology to simultaneous multi-band imaging instruments, where only a limited number of wavelength channels are available. Such an application is particularly attractive because these instruments are commonly used for time-domain studies and typically acquire sequences of short-exposure observations, often with integration times of only a few seconds or even below one second. In this context, the method could provide not only estimates of atmospheric and dome-induced turbulence parameters but also valuable information on their temporal evolution, thereby enabling direct monitoring of turbulence variability from routine scientific observations.

\section{Applications to PSF reconstruction and AGN--host deblending}

Beyond turbulence characterization itself, these diagnostics can support the construction of wavelength-dependent PSF models for seeing-limited observations. Since the IQ at the focal plane is determined by the prevailing atmospheric and dome-turbulence conditions, estimates of $\epsilon_0$, \LO, and the dome-turbulence contribution can be used to predict the evolution of the PSF across the spectral range of an observation.

This approach is particularly relevant for the analysis of type-I active galactic nuclei (AGN), where the unresolved nuclear emission contaminates the spectra of the host galaxy. In previous work, we showed that the wavelength dependence of the PSF can be exploited to construct a three-dimensional PSF model and separate the AGN contribution from the surrounding host-galaxy emission in seeing-limited MUSE observations  ([\citenum{2025Garcia-Lorenzo}]). The method uses PSF measurements obtained from spectral regions dominated by the broad AGN emission lines and extends them across the full spectral range using the expected wavelength dependence of the IQ derived from atmospheric turbulence theory. The resulting three-dimensional PSF can then be scaled to the AGN spectrum and subtracted from the data cube, enabling an accurate analysis of the circumnuclear stellar and gaseous components. 

The turbulence characterization techniques presented in this work provide a natural framework for improving PSF modelling in seeing-limited observations. The recovered turbulence parameters can be used to constrain physically motivated wavelength-dependent PSF models that are directly linked to the atmospheric and enclosure conditions prevailing during the observations. Furthermore, when independent measurements of the relevant turbulence parameters are available from dedicated site-monitoring instruments, reliable PSF models may be constructed even in the absence of suitable reference stars or direct PSF measurements within the science field. This capability may be particularly valuable for AGN–host deblending, image reconstruction, and future PSF-aware analyses of seeing-limited spectroscopic observations. In the longer term, such turbulence-informed PSF models could provide a complementary route to PSF reconstruction in seeing-limited observations, bridging site-monitoring measurements and science data analysis.

\section{Conclusions}

In this work, we have explored the use of focal-plane information in seeing-limited observations as a diagnostic of atmospheric and dome-induced turbulence. Building on the methodology recently developed for IFS, we investigated its applicability to other observing modes and discussed several scientific applications of the recovered turbulence parameters.

Our results show that the variation of IQ with wavelength is sensitive to atmospheric seeing and outer scale, as well as to dome-induced turbulence, whose behaviour may depart from the Kolmogorov prediction. The proposed methodology therefore provides a promising observational probe of both atmospheric and local turbulence properties in routine scientific observations.

We have shown that the methodology originally developed for IFS observations can be extended to other observing modes that provide spatial information at multiple wavelengths under homogeneous observing conditions. As a first exploratory test, long-slit spectroscopic observations obtained with OSIRIS at the GTC yield turbulence parameters compatible with typical atmospheric conditions at the Observatorio del Roque de los Muchachos, supporting the potential applicability of the method to spectroscopic observations beyond IFS data. 

In addition, by emulating a five-channel simultaneous broadband imager using MUSE data cubes, we find that turbulence parameters derived from only five simultaneous wavelength measurements remain consistent, within the current uncertainties, with those obtained from the full 30 narrow-band spectral sampling. This agreement demonstrates that a relatively small number of simultaneous wavelength channels can provide meaningful constraints on atmospheric and dome-induced turbulence. Because such instruments are widely used in time-domain astronomy and frequently operate with integration times of only a few seconds, or even at sub-second cadence, they offer the possibility of studying the temporal variability of atmospheric and dome-induced turbulence directly from routine scientific observations. 

Beyond turbulence characterization itself, the recovered parameters provide physically motivated constraints for wavelength-dependent PSF models. Such models can improve PSF reconstruction, AGN–host deblending, image reconstruction, and other PSF-sensitive analyses. In the longer term, turbulence-informed PSF models may offer a complementary link between site-monitoring measurements and the scientific exploitation of seeing-limited observations, enabling more realistic descriptions of image formation under varying atmospheric conditions.

At the same time, the present work should be regarded as an exploratory study. The atmospheric and dome-turbulence parameters recovered from the focal-plane data are generally consistent with values reported for the corresponding observatories, but a rigorous validation will require dedicated comparisons with independent measurements of atmospheric outer scale and dome seeing obtained simultaneously with the scientific observations. Such validation will be particularly important for assessing parameter degeneracies, refining dome-seeing models, and establishing the accuracy and operational applicability of the method under a wider range of observing conditions.

Overall, these results demonstrate that standard scientific observations contain valuable information on atmospheric and dome-induced turbulence that can complement dedicated site-monitoring instrumentation. The methodology presented here could potentially be incorporated into routine observatory pipelines, enabling turbulence diagnostics from standard science observations while providing a valuable complement to existing monitoring systems. More broadly, this approach offers a promising route toward improved turbulence characterization, PSF modelling, image-quality prediction, and the optimization of future ELT operations.

\acknowledgments 
 
Based on data obtained from the ESO Science Archive Facility (DOI: https://doi.eso.org/10.18727/archive/41). We thank Antonio Cabrera-Lavers for providing long-slit spectrophotometric standard-star observations obtained with OSIRIS@GTC for the exploratory analysis of the methodology on this type of data. The authors acknowledge support from the Spanish Ministry of Science and Innovation through the Spanish State Research Agency (AEI-MCINN/10.13039/501100011033) via the grants “Participation of the Instituto de Astrofísica de Canarias in the development of HARMONI: D1 and Delta-D1 phases” (PID2019-107010GB-100), “Participation of the Instituto de Astrofísica de Canarias in the development of HARMONI: Delta-D1 phase and Rescope Study” (PID2022-140483NB-C21 and PID2024-158231NB-C21). The authors also acknowledge support from the project “IPOICE: Apoyo al desarrollo en el IAC de la preóptica (IPO) y el control electrónico (ICE) de HARMONI”, funded by ESO-MICIN and the European Union NextGenerationEU/RTRP (Agreement No. 64365/ESO/15/66976/JSC). The authors acknowledge the use of ChatGPT (OpenAI) for assistance with language editing and improving the clarity of the manuscript.

\bibliography{report} 
\bibliographystyle{spiebib} 

\end{document}